\documentclass[epj]{svjour}
\usepackage{graphicx}
\usepackage{amsmath}
\usepackage{amssymb}

\begin{document}

\title{Quantum computer with cold ions in the Aubry pinned phase}

\author{
D.L.Shepelyansky}

\institute{
Laboratoire de Physique Th\'eorique, IRSAMC, 
Universit\'e de Toulouse, CNRS, UPS, 31062 Toulouse, France
}

\titlerunning{Quantum computer with cold ions in the Aubry phase}
\authorrunning{D.L.~Shepelyansky}

%\date{\today}
\date{Dated: 26 February 2019}

\abstract{It is proposed to modify the Cirac-Zoller
proposal of quantum computer with cold ions in
a global oscillator trap potential
by adding a periodic potential
with an incommensurate average ratio
of number of ions to number of periods being
order of unity.
With the increase of 
the periodic potential amplitude the system
enters in the Aubry pinned phase
characterized by quasi-frozen
positions of ions and a gap
of their first phonon excitations becomes
independent of number of ions.
This gives hopes that this quantum computer
will be really scalable.
It is argued that the usual
single- and two-qubit gates
can be realized between
the nearby ions in the Aubry phase.
The possibilities of experimental
realizations of a periodic
potential with microtrap arrays
or optical lattices
are discussed. It is pointed that the disorder
of distances between microtraps with one ion per trap
can lead to the Anderson localization
of phonon modes with interesting possibilities
for ion quantum computing.
}

%% PACS to be updated or removed
%\PACS{
%{89.75.Fb}{
%Structures and organization in complex systems}
%\and
%{89.75.Hc}{
%Networks and genealogical trees}
%\and
%{89.20.Hh}{
%World Wide Web, Internet}
%}

\maketitle

\section{Introduction}
\label{sec:1}

The creation of a scalable quantum computer for 
generic computational tasks is an important challenge of 
modern quantum technology \cite{nielsen}.
One of the first physical proposals
of such a computer
is the Cirac-Zoller quantum computer of 1995
with a chain of cold ions placed 
in an oscillator trap potential \cite{zoller}.
Indeed, at that time the storage of cold ions 
already allowed to keep several tens of ions in 
a storage ring \cite{walther}.
Thus soon after the proposal a two-qubit 
gate with a conditioned phase shift
had been realized \cite{monroe1995}
followed later by realization
of a few other two-qubit gates 
\cite{sackett2000,demarco2002,leibfried2003}.
Simple quantum algorithms \cite{gulde2003}, 
a set of universal gates with two ions \cite{schmidt2003}
and a creation of various entangled states \cite{roos2004}
had been also reported.
The experimental progress with cold ion experiments
is reviewed in \cite{blatt2008,wunderlich,blatt2012,wineland2013,ionrevmit}.
At present up to 100 ions can be routinely trapped
for hours in a linear trap configuration \cite{monroe2018}.
Recently various ionic quantum computations has been performed with
up to 11 qubits \cite{kim10q,monroe2019nat,11q,monroefiedtheo,roos2019sci}.
This experimental progress makes cold ions
to be very attractive for scalable quantum computer realization.
Their important physical advantages are related
to possibilities of individual addressing of a selected ion
by a laser beam and low temperatures reached  experimentally.
Since up to 11 qubits are now used in the ionic quantum computations
it becomes of primary importance to have the firm
concept of scalable ionic quantum computer.

However, the scalable quantum computation
with ion-trap computers is not so easy to reach
even if about 100 ions can be now trapped for hours.
Thus, the original Cirac-Zoller
proposal \cite{zoller}
is not really scalable for a very large number of ions.
Indeed, the coupling between ion chain and the internal
ion levels decreases with the number $N$ of trapped ions 
as $1/\sqrt{N}$ (see Eq.(1) in \cite{zoller} ).
 Also, the ion chain oscillation
frequency $\omega_{tr}$ is unavoidably decreasing if 
the number of ions in the trap is
growing with a constant average distance between ions. 
Thus the gap between the ground state
and the first excitation
of ion chain drops with $N$.
It is proposed to avoid these problems
with a modular type architecture 
with quasi-separated groups of ions
with a further adiabatic transfer of 
quantum information between groups.
However, the practical realization
of this concept is not an easy task.

Here I propose another concept of 
quantum computer with cold ions in
a linear configuration
based on the Aubry pinned phase \cite{aubry}.
In this proposal the linear chain of
ions is placed in a periodic potential (or lattice),
created by external fields,
and a global oscillator trap potential.
It is assumed that there is
an incommensurate density of ions
$\nu =N/L \sim 1.618$
(ratio of number of ions $N$ per number of 
potential periods $L$).
In the limit of small potential amplitude
the system is reduced to the Cirac-Zoller proposal.
In this regime the spectrum 
of ion excitations have an 
almost acoustic spectrum  
starting from  $\omega_{tr}$ 
which goes to zero in the limit of large
number of ions. However, when 
the lattice amplitude $K$ becomes larger than 
a certain critical value $K_c$
the chain enters in the Aubry pinned phase
with the appearance of  optical gap
$\omega_g$ of excitations being independent of the chain length 
and the number of ions placed in it.
The physics of this transition
is related to the dynamical symplectic maps,
invariant Kolmogorov-Arnold-Moser (KAM) $\;\;\;$ curves
and the fractal cantori replacing these
curves above the transition to the Aubry pinned
phase corresponding to the chao\-tic map dynamics.
Since the spectral gap $\omega_g$ is independent
of the system size it is possible
to place unlimited number of ions in such a system.

The first analytical and numerical studies of ions in a periodic potential
had been done in \cite{fki2007} where its
physical properties and the critical point
of Aubry had been determined. 
The cold ion experiments  had been started
in \cite{haffner2011} and the signatures of the 
predicted Aubry transition have been reported
recently by the Vuletic group \cite{vuletic2015sci,vuletic2016natmat}
with up to 5 ions. The Aubry phase with
chains of larger number of ions 
is under investigations \cite{ions2017natcom}.
Recently, the transport properties 
of charges in a periodic 1D and 2D lattices
have studied analytically and numerically
in \cite{diode,ztfki}. However, in these studies \cite{fki2007,diode,ztfki}
ions or charges were considered 
without internal states while 
they are essential since they form
a qubit for a given ion
and the interactions between 
internal ion states (usually S and D states 
are used that give a qubit lifetime of about a second \cite{zoller,blatt2012}).
Also the coupling between internal ion states
and  spacial motion
of ions is essential for the realization
of universal quantum gates.
These features are discussed in this work
with arguments about the advantages
of ions placed in a lattice
of Aubry phase.

The paper is constructed as follows:
the system description and its physical properties are
given in Section~\ref{sec:2}, the quantum gates 
with ions in the Aubry phase are discussed in Section~\ref{sec:3}
and the discussion of the results and possible experimental realizations
are given in Section~\ref{sec:4}.

\section{System description and properties}
\label{sec:2}

The motion of ions in a periodic potential and a
global oscillator potential is described by the Hamiltonian \cite{fki2007}:
\begin{equation}
H = \sum_{i=1}^{N} (\frac{P_i^2}{2}+\frac{{\omega_{tr}}^2}{2} x_i^2 - K \cos x_i) +
\sum_{i > j}\frac{1}{|x_i-x_j|}  \; .
\label{eq1}
\end{equation}
Here $P_i, x_i$ are ion momentum and position, $K$ gives the amplitude
of periodic potential and all $N$ ions are placed in a harmonic trap
potential with frequency $\omega_{tr}$. The Hamiltonian
is written in dimensionless units
where the potential period is $\ell=2\pi$
and ion mass and charge are $m=e=1$.
In these atomic-type units 
the physical system parameters are expressed in
units:   $r_a= \ell/2\pi$ for length,
$\epsilon_a = e^2/r_a = 2\pi e^2/\ell$ for energy,
$E_{adc} = \epsilon_a/e r_a$  
for applied static electric field,
$v_a=\sqrt{\epsilon_a/m}$ for particle velocity,
$t_a =  e r_a \sqrt{m/\epsilon_a}$ for time $t$.

The physical properties of this system
has been analyzed in detail in \cite{fki2007}.
They are not sensible to the boundary conditions
so that instead of global oscillator
potential one can consider the ion chin with
fixed ends or hard wall boundary conditions
\cite{ztfki,ztzs}.

The equilibrium positions of ions are
determined by the condition $P_i=0$
and $\partial H/\partial x_i = 0$. 
In the approximation of interactions only
between nearest neighbors this give the 
recursive map for equilibrium ion positions $x_i$:
\begin{eqnarray}
p_{i+1} = p_i + K g(x_i) \; , \; \; x_{i+1} = x_i+1/\sqrt{p_{i+1}} \; .
\label{eq2}
\end{eqnarray}
Here $p_i = 1/(x_{i}-x_{i-1})^2$ is 
the effective momentum conjugated to $x_i$
and the kick function is $K g(x)=-\omega^2 x - K \sin x$.
The numerical simulations performed in
\cite{fki2007,diode,ztfki,ztzs} confirm
that this approximation provides a good
description of real ion positions obtained 
by numerical simulations. Thus the nearest 
neighbor interactions between ions are dominant.

The map description (\ref{eq2})
provides important links
with the generic properties of dynamical symplectic maps
(see e.g. \cite{chirikov,lichtenberg,meiss}). 
The equation for $x_{i+1}$ 
can be locally linearized in $p_{i+1}$
near the resonant values of $p_r \approx 2\pi/\nu$
defined by the condition $x_{i+1} =  x_i + 2\pi m$
where $m$ are integers 
(see examples in \cite{chirikov,lichtenberg}).
This leads to the local description of dynamics 
by the Chirikov stadard map \cite{fki2007}:
\begin{eqnarray}
y_{i+1} = y_i - K_{eff} \sin x_i \; , \; \; x_{i+1} = x_i - y_{i+1} \; ,
\label{eq3}
\end{eqnarray}
where $y_i=\alpha (p_i - p_r)$, 
$\alpha =  1/(2 {p_r}^{3/2}) = (2\pi/\nu)^3/2$
and the dimensionless chaos parameter 
$K_{eff}=\alpha K = K(2\pi/\nu)^3/2$.

%
%%\begin{figure}[t!]
%%\epsfxsize=3.2in
%\epsfysize=2.6in
%\hskip 1.9 cm \epsffile{fig1.ps}
%%\epsffile{fig1.eps}
%\includegraphics[width=.9\linewidth]{fig2.ps}
%%\vglue +1.5cm
\begin{figure}[!th]
\centering
\includegraphics[width=\columnwidth]{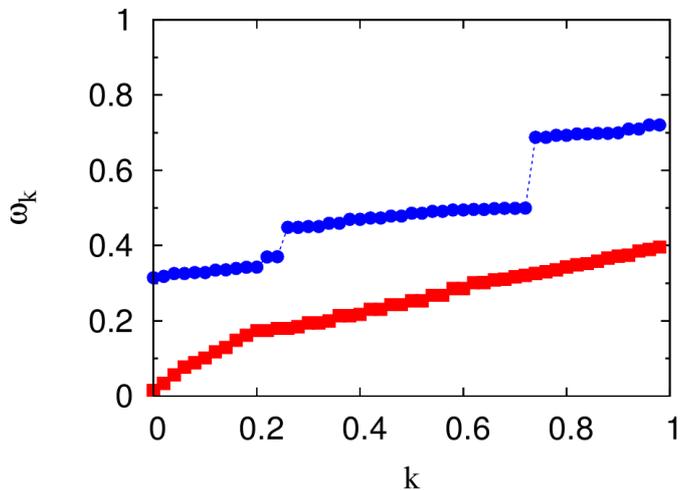}
\caption{
Phonon spectrum $\omega(k)$ as a function of
scaled mode number $k=i/N$ ($i=0,\ldots,N-1$)
for the KAM sliding phase at $K=0.03$ (bottom curve, red squares) 
and the Aubry pinned phase at 
$K=0.2$ (top curve, blue points) for 
$N=50$ ions in a trap with frequency $\omega_{tr}=0.014$
which approximately gives the golden mean
density in the central $1/3$ part of the chain
(after \cite{fki2007}).
}
\label{fig1}       
\end{figure}

This local description
corresponds to the linear-spring forces
locally acting between particles that in fact represents the
Frenkel-Kontorova model describing commensurate-incommensurate
transition in solid states systems \cite{obraun}. 
Thus the properties of  this system of ions in a
periodic potential can be understood from the properties of the Chirikov 
standard map which describes the local 
dynamics of various physical systems (see e.g. \cite{stmapscholar}).

%%\begin{figure}[t!]
%%\epsfxsize=3.2in
%\epsfysize=2.6in
%\hskip 1.6 cm \epsffile{fig2.ps}
%%\epsffile{fig2.eps}
%%\vglue +1.3cm
\begin{figure}[!th]
\centering
\includegraphics[width=\columnwidth]{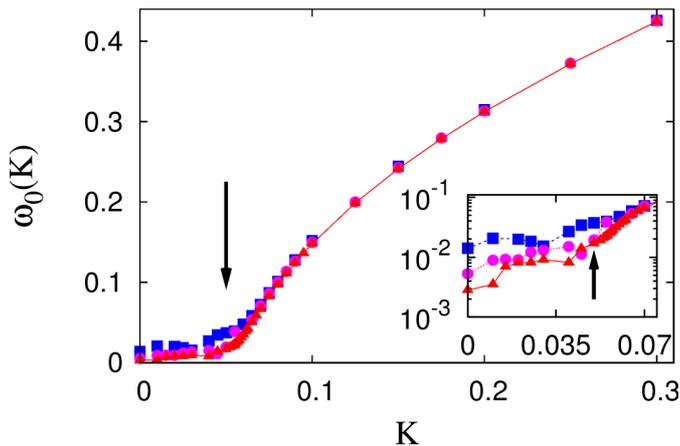}
\caption{(Color online) Minimal excitation frequency
$\omega_0(K)$ as a function of periodic potential strength
$K$ for the golden mean ion density $\nu_g = 1.618...$ and 
number of ions $N=50$ (blue squares; $\omega_{tr}=0.014$),
$N=150$ (magenta circles; $\omega_{tr}=0.00528$),
$N=300$ (red triangles, $\omega_{tr}=0.00281$).
The critical point  $K_c \approx 0.05$ of Aubry transition
is marked by arrow;
inset shows data near $K_c$ (after \cite{fki2007}).
}
\label{fig2}       
\end{figure}

At small $K$ or $K_{eff}$ the phase space 
of maps (\ref{eq2}) and  (\ref{eq3})
is covered by
the invariant KAM curves characterized
by irrational rotation number $r= <x_i - x_0>/2\pi i = \nu$
which gives an average distance (phase) between ions
related to the average ion density $\nu$.
The oscillations of ions near the equilibrium positions
has the acoustic excitation spectrum
$\omega_k \approx C_v k + \omega_{tr}$
where $k=i/N$ plays the role of wavevector number
and $C_v \sim 1$ is the sound velocity.

For the  Chirikov standard
map the last invariant curve 
with the golden mean rotation number 
$r_g=\nu_g=1.618...$ is destroyed at $K_{eff} \approx 1$ 
leading to a global chaos and diffusion in $y$ \cite{chirikov}.
For the case of ions with density $\nu$ this gives 
the critical amplitude of potential \cite{fki2007}:
\begin{equation}
K_c(\nu) \approx 0.034 (\nu/\nu_g)^3 \;, \;\;  \nu_g = 1.618... \; .
\label{eq4}
\end{equation}
This theoretical dependence is recently confirmed 
by extensive numerical simulations \cite{ztfki}.
For $\nu=\nu_g$ the numerical results give
$K_c =0.0462 $ \cite{fki2007,ztfki,ztzs}
that is slightly higher than the theoretical value 
due to the local approximation used in the reduction
to the Chirikov standard map.

For $K>K_c(\nu)$ the invariant KAM curve
is destroyed and it is replaced by a fractal cantori invariant set
as proved by Aubry in \cite{aubry}. The configuration of particles
corresponding to this invariant set has the minimal energy
and thus represents the ground state of the system.
The spectrum of ion oscillations near these 
ground state positions is characterized by the optical gap
$\omega_g \sim \sqrt{K}$. Thus in difference
from the KAM sliding phase at $K<K_c$
for $K>K_c$ we have the Aubry pinned phase 
where the ion chain is pinned by the lattice.

The example of excitation spectrum for the KAM and Aubry 
phases is shown in Fig.~\ref{fig1} taken from \cite{fki2007}.
The dependence of the minimal excitation frequency
$\omega_0(K)$ on potential amplitude $K$ is shown 
in Fig.~\ref{fig2} taken from \cite{fki2007}.
For these data the trap frequency $\omega_{tr}$ is
chosen in such a way that, at a given number of ions $N$ in the trap,
the central $1/3$ part of the chain keeps the fixed
density $\nu \approx 1.618$ when the number of ions $N$ is growing.
Due to this condition at $K=0$, corresponding to the Cirac-Zoller
proposal \cite{zoller}, the trap becomes more and more soft
and $\omega_{tr} \sim 1/\sqrt{N} \rightarrow 0$.
Indeed, we want to keep the distance between ions
in the center to be independent of $N$
and thus size of the chain $x_{chain} \sim  N/\nu$ is growing 
since it is approximately determined by the condition
at the chain end $F_{chain} \sim {\omega_{tr}}^2  x_{chain} \sim \nu^2$
that gives the above dependence  $\omega_{tr} \sim 1/\sqrt{N}$. 

Thus for $K < K_c$ the lowest excitation frequency 
goes to zero with the increase of number of ions in the trap.
Hence the Cirac-Zoller proposal in not really scalable.
In contrast for $K > K_c$ the lowest frequency excitation
in independent of $N$ as it is well seen in Fig.~\ref{fig2}.
Thus this Aubry pinned phase has certain chances
to represent a scalable architecture for a
quantum computer with cold ions.

Indeed, for the quantum case
the energy of lowest phonon excitation is
$E_0 = \hbar \omega_0(K) = \hbar \omega_g$ being independent of $N$.
For a temperature $T \ll  \hbar \omega_0(K)$
the phonon excitations become frozen
and should not perturb the accuracy of quantum gates operations. 

There are also another type of quantum excitations
in the quantum ion chain inside the Aubry pinned phase.
In fact the Aubry theorem \cite{aubry},
which guaranties that the Aubry cantori ground state
has the minimal energy $E_A$ of the classical ion chain
is mathematically correct but 
it is wrong from the physical view point.
Indeed, in the classical chain
there are exponentially many static configurations 
of  ions which number $N_s$ grows exponentially 
with the number of trapped ions $N$.
In addition the energies of these configurations 
are approaching exponentially close to the Aubry
ground state energy $E_A$ with increase of $N$
(see Fig.~4 in \cite{fki2007} where this feature
is clearly illustrated). In fact this property
is similar to the random spin glass systems \cite{parisi}.
However, in our case the randomness is absent 
and the system is described by a rather simple 
deterministic Hamiltonian (\ref{eq1}).
Thus the Aubry pinned phase
represents the dynamical spin glass system
with an enormous amount of quasi-degenerate 
configurations in a vicinity
of the Aubry ground state.

In the quantum case there is 
quantum tunneling between these quasi-degenerate
configurations that can be viewed as instanton
excitations. However, for small dimensionless Planck constant $\hbar_{eff}$
the gas of instantons is very dilute
and the tunneling times are enormously long \cite{fki2007}.
Thus on a scale of typical tunneling time
$t_{tul} \propto \exp(A/\hbar_{eff})$ we can consider the
ions to be frozen at their positions
(here $A \propto K$ is a typical
action between energy minima coupled by tunneling). 
The dimensionless Planck constant is
$\hbar_{\rm eff}=$ $\hbar/$ $(e \sqrt{m \ell/2\pi})$ and
for a typical lattice period $\ell \approx 1 \mu m$,
ion density $\nu \sim 1$ and ${^{40}}\rm Ca^{+}$ ions
we have very small $\hbar_{\rm eff} \approx  10^{-5}$.
Thus the quantum ions can be considered as frozen
at their configuration positions for the whole time scale 
of quantum computations.

\section{Quantum gates}
\label{sec:3}

As in the proposal of Cirac-Zoller \cite{zoller}
I assume that the qubit is formed by
two internal levels $S_{1/2}$ and  $D_{5/2}$
of ${^{40}}\rm Ca^{+}$ ion with a radiative life-time of 
more than one second. All single-qubit gates
can be realized by laser pulses as described in
\cite{zoller,blatt2012}. 
At present  these gates are routinely 
performed with the fidelity exceeding 0.99 \cite{blatt2012}.
The individual accessing of ions is also
available in experiments with ion spacing
of about $5 \mu m$ \cite{blatt2012}.

Since single-qubit gates with ions are reliable
the most important for quantum computations become two-qubit gates
which in combination with single-qubit gates allow
to perform universal quantum computations \cite{nielsen}. 
There are three types of two-qubit gates
usually discussed for cold ions 
(see e.g. review \cite{blatt2012}): the Cirac-Zoller gate \cite{zoller},
the Molmer-Sorensen gate \cite{molmer} and
the geometric phase gate \cite{leibfried2003}
closely related to the Molmer-Sorensen gate.

In all these gates the motional oscillator states of ions (sideband)
with frequency $\omega_0$
are coupled by a tuned laser pulse with internal $S-D$
levels of ions. Usually as an example one considers
two ions with two internal levels and their sideband modes \cite{blatt2012}.
The laser pulse duration is selected in a way 
allowing to perform two-qubit gate.
In the case of long ion chain in an oscillator trap 
the operational frequency of the Cirac-Zoller gate 
is proportional to the strength of coupling
of internal levels with the whole chain oscillator state (the bus mode)
which decreases with the number of ions as $1/\sqrt{N}$.

Another possibility for qubit is to use, instead of $S-D$ levels,
the hyperfine-split 
$^2S_{1/2}$ ground level with an energy difference of $12.64$ GHz
with the life-time of 1.5s, as it is done in \cite{monroe2019nat}.

For the Molmer-Sorensen gate both ions are irradiated with a 
bichromatic laser field with frequencies 
$\omega_0 \pm (\omega_{qubit} + \delta)$ 
tuned close to the red and the blue sideband of a collective mode
(see Fig.14 in \cite{blatt2012}). This approach allowed to 
create experimentally Bell states with a fidelity 99.3\% \cite{blatt2012}

The same gates can be implemented for ions in the Aubry pinned phase.
In this case the interaction of ions is well approximated 
by the nearest neighbor interactions as 
is discussed in the previous Section
with the map (\ref{eq2}) description of 
equilibrium ion configurations.
The oscillations of ions in a vicinity
of equilibrium positions is harmonic 
and we can consider them as  sideband
transitions for laser pulses as 
for the two-qubit gates considered above.
Since the interactions are dominated by nearest neighbors
the coupling between internal qubit levels 
and ion oscillator mode is independent of
the number of ions in the chain.
The frequency of this oscillator or phonon mode gap
is $\omega_g = \omega_0(K)$ 
being also independent of the chain length as it is 
shown in Fig.~\ref{fig2}.

The construction of two-qubit gates should also take into account that 
when cold ions are cooled and loaded in the Aubry pinned phase
it is most probable that they will be located in one
of quasi-degenerate static configurations.
Thus the distances between nearby ions
will be somehow irregular
that will affect the interactions between 
specific pairs of ions. However, it is
possible to determine experimentally the actual ion
positions and then to adapt the laser pulses of two-qubit
gates to these experimentally found  ion positions.
In a sense for a good work of a piano
each string should be checked and 
adapted. Here, for quantum gates 
with ions in the Aubry phase 
we have a similar situation.

However, there are certain points that require
additional investigations. The low energy
phonon exciations with the lowest phonon
frequency $\omega_0(K)$ are excited 
by a tuned laser pulse which acts
mainly on one or two nearby ions.
Thus there is a question how this excitation
will propagate along the chain of 
ions in the Aubry pinned phase.
This propagation or spreading along the chain
depends on two main factors:
the localization properties of phonon 
modes in the pinned phase
and the rate of decomposition
of local ion oscillations into these photon modes.
At present very little analysis has been performed
for these important properties of ionic
phonon modes in the Aubry phase.
Examples of a few phonon eigenmodes
are given in \cite{fki2007} (see Figs.9,10 there).
Some of modes look to be localized some of them
have spreading over several ions.
The spreading rate of one or two ion oscillations
has not been studied and require further investigations.

Thus there are open questions on
the possible fidelity and accuracy
of two-qubit gates for cold ions in 
the Aubry phase. 
 
Finally, it is important to note 
other proposal \cite{zoller2000nat}
where ions are assumed to be placed 
in an array of equidistant microtraps
in 1D (and even in 2D). Formally in 1D
this approximately corresponds to
the case of periodic potential
considered here at the filling factor
$\nu =1$ (one ion per period)
with a sufficiently high barrier.
In this proposal the two-qubit gates are again
constructed assuming the harmonic 
approximation of ion motion inside the 
minitraps. However, at $\nu=1$ we have 
a periodic structure of ions and according to
the Bloch theorem the spectrum of
ionic phonon oscillations near equilibrium
positions in the minima of periodic potential
will correspond to a ballistic propagation 
of waves along the ion chain that
will destroy the local oscillator approximation
used in the derivation of quantum gates.
In contrast for irrational filling factor
$\nu=1.618...$ discussed above the phonon modes
will see an incommensurate potential 
with a possibility of the Aubry-Andre localization
of the ionic phonon modes
(the Aubry-Andre transition in an incommensurate potential
is found in \cite{aubryandre}
and observed in cold atom experiments \cite{inguscio,bloch}).
As pointed above, there are some signatures of localization
of ionic phonon modes shown in Figs.9,10 in \cite{fki2007}
but a much more detailed analysis of these modes
and disintegration of initial excitation
of a specific ion oscillations with time are required.

There had been certain attempts to 
study ionic phonon modes
(see e.g. \cite{morogi2016pra})
but the direct connection with the KAM - Aubry transition
in the related symplectic maps had not been used 
without which it is rather difficult to understand the 
properties of these nonlinear strongly interacting many-body 
systems.
The proposals of using an anharmonic
linear ion trap to obtain a scalable trap \cite{lin2009}
also do not present deep analysis of spectrum of 
phonon modes and the spreading of excitation
of a specific ion (e.g. the central part of the
chain proposed there is approximately homogeneous and 
has the same problems of localization of phonon modes).

The proposals to study 2D ion systems
\cite{zoller2000nat,blatt2d,schaetz2d} are also facing the problem of
understanding of the Aubry transition in 2D. In addition to that
the problem of phonon spectrum and properties of phonon
modes in 2D is much more involved comparing to 1D case.
However, the recent results for charge transport
of Wigner crystal in 2D periodic potential \cite{diode}
allows to hope that under certain conditions 1D results
can be extended to 2D case. 

\section{Discussion}
\label{sec:4}

In this work I analyzed the properties of 
cold ion chain  in a periodic potential
which amplitude locates ions inside the Aubry pinned phase.
The emergence of Aubry
transition from KAM sliding phase
to Aubry pinned phase takes place when the potential exceeds
a critical value $V_{A}=K_c(\nu)e^2/(\ell/2\pi)$.
For a typical lattice period 
$\ell=1\rm\mu m$ and dimensionless ion density per period
$\nu = 1.618$ this corresponds to
$\;\;\;\; V_A \approx 3 k_B Kelvin$. Apparently this 
amplitude significantly exceeds the amplitudes
reachable with presently available laser power for optical lattices.
Usually the optical lattice amplitude
is assumed to be able to  reach  the value 
$ V_{A} \approx \- 10^{-3} k_B Kelvin$ 
(see e.g. review \cite{schaetz}).
However, in recent experiments \cite{drewsen},
published after the submission of this work,
it was possible to reach $ V_{A} \approx 0.025 k_B Kelvin$
with $\ell \approx 20 \rm\mu m$.
The recent result presented in \cite{ztfki}
shows that the border of the Aubry transition 
drops as a cub of density $\nu$ (\ref{eq4})
so that at these values of $ V_{A}$, $\ell$ 
it is possible ion chain in the Aubry pinned phase at
$\nu \approx 0.618$. It should be noted that
in the optical lattice the
qubit state $D$ may be not affected by an optical
potential which may be generated   e.g. by $S-P$
transition. In this case for the first
experimental realisations of two-qubit quantum gates
it is possible 
to use the hyperfine-split $^2S_{1/2}$ ground level
as it is done in \cite{monroe2019nat}. 

In contrast to optical lattices the radio-frequency (RF) traps
provide the potential depth $V_{RF} \approx 10^4 k_B Kelvin$
that is significantly above the estimated Aubry transition
potential amplitude \cite{schaetz}. 
At present there is a significant miniaturization
of these RF traps with sizes going down to tens of microns \cite{schaetz2d,schaetz}.
Thus such microtrap linear arrays can model
the periodic potential considered here
with  high amplitudes of periodic potential 
allowing to place ions in the Aubry pinned phase.
There is also progress with the Penning mircotraps
of about 10 micron size \cite{home}. 
In principle these traps have a 2D potential minimum
but we can home that one of these two directions may be designed
to be significantly larger than other. Thus with 
orientation of axis with lowest frequency along the 
ion chain direction one can realize a quasi-one-dimensional 
situation with an effective 1D periodic potential discussed here.
We note that in such traps both $S$ and $D$ states feel the periodic 
microtrap potential. 
Thus the linear array of RF or Pinning microtraps of such type
would allow to observe the Aubry transition
and hopefully to perform 
scalable quantum computations 
with cold ions in the Aubry pinned phase.
 
The important message of this work is
that in the Aubry pinned phase
there is a gap for energy excitations
independent of number of ions in a linear
configuration considered here.
This is the good feature of this Aubry phase
but still there are open questions to be resolved.
Indeed, in the limit of large system size the 
spectrum of ionic phonons is dense
so that some phonons inevitably
have very close frequencies.
However, the question if these modes are coupled
or not is not so simple.
For example, for the Aubry-Andre model \cite{aubryandre}
the spectrum of modes is dense since the system
size is infinite but the modes are exponentially localized
and thus there are practically no interactions
between modes localized far from each other.
In the Aubry pinned phase we may hope
that there will be just such a situation.
But in a difference from the linear case of
the Aubry-Andre model we have nonlinear couplings between
phonons in the Aubry pinned phase of our model (\ref{eq1}) and the detailed 
analysis of these nonlinear phonon interactions 
should be performed in detail in the further studies.
Due to the long range of Coulomb interaction between ions
the investigation of properties of these 
ionic phonon modes will be necessary for any
scalable realization of ion quantum computer.

Finally, it is interesting to note
another regime of ion microtrap arrays
which has certain similarities
with the Aubry-Andre localization \cite{aubryandre}.
Indeed,  it is possible to place the microtraps 
with a random distance between each pair of
nearby traps (e.g. an average distance
between traps is $20 \rm \mu m$ 
and for each nearby traps the actual
distance changes randomly in the range
between $15 \rm \mu m$ and $25 \rm \mu m$).
It is known that in disordered systems
the Anderson localization of modes
can take place \cite{anderson1958}.
Moreover, in the thermodynamical limit
all eigenmodes are exponentially
localized in 1D random potential 
\cite{pastur}. 
Due to a finite minimal/maximal distance between 
microtraps this system is also characterized by
the finite gap of excitations independent of the number of ions
in such a linear chain.
Thus it may be also important
to investigate the ion quantum computer 
in the Anderson localized phase
created by disorder of distances 
between microtraps with one ion per trap.

% Acknowledgments before appendices 
%-------------------------------
\section{ Acknowledgments}
I thank P.Zoller for discussion and pointing to me 
Refs.~\cite{schaetz,home}.
This work was supported in part by the Pogramme Investissements
d'Avenir ANR-11-IDEX-0002-02, reference ANR-10-LABX-0037-NEXT France
(project THETRACOM).


\begin{thebibliography}{99}

\bibitem{nielsen} M.A.~Nielsen and I.~Chuang, 
        {\it Quantum computation and quantum information},
        Cambridge University Press, Cambridge UK (2000).
\bibitem{zoller} J.I.~Cirac and P.~Zoller,
         {\it Quantum computations with cold trapped ions},
         Phys. Rev. Lett. {\bf 74}, 4091 (1995).
\bibitem{walther} G.~Birkl, S.~Kassner, and H.~Walther,
         {\it Multiple-shell structures of
          laser-cooled $^{24}{Mg}^+$ ions in a quandrupole storage ring},
          Nature {\bf 357}, 310 (1992).
\bibitem{monroe1995} C.~Monroe, D.M.~Meekhof, B.E.~King, W.N.~Itano, and D.J.~Wineland, 
         {\it  Demonstration of a fundamental quantum logic gate},
         Phys. Rev. Lett. {\bf 75}, 4714 (1995).
\bibitem{sackett2000} C.A.~Sackett, D.~Kielpinski, B.E.~King, C.~Langer, V.~Meyer, 
         C.J.~Myatt, M.~Rowe, Q.A.~Turchette, W.M.~Itano, D.J.~Wineland, and C.~Monroe, 
         {\it Experimental entanglement of four particles},
         Nature {\bf 404}, 256 (2000).
\bibitem{demarco2002} B.~DeMarco, A.~Ben-Kish, D.~Leibfried, V.~Meyer, 
         M.~Rowe, B.M.~Jelenkovic, W.M.~Itano, J.~Britton, C.~Langer, 
         T.~Rosenband, and D.J.~Wineland, 
          {\it Experimental demonstration of a controlled-NOT wave-packet gate},
         Phys. Rev. Lett. {\bf 89}, 267901 (2002).
\bibitem{leibfried2003} D.~Leibfried, B.~DeMarco, V.~Meyer, 
            D.~Lucas, M.~Barrett, J.~Britton, 
          W.M.~Itano, B.~Jelenkovic, C.~Langer, T.~Rosenband, and D.J.~Wineland, 
          {\it  Experimental demonstration of a robust, high-fidelity geometric 
          two ion-qubit phase gate},
           Nature {\bf 422}, 412 (2003).
\bibitem{gulde2003} S.~Gulde,  H.~Haffner, M.~Riebe, G.~Lancaster, C.~Becher,  
         J.~Eschner, F.~Schmidt-Kaler, I.L.~Chuang, and R.~Blatt, 
         {\it Quantum information processing with trapped $Ca^+$ ions},
         Proc. R. Soc. Lond. A {\bf 361}, 1363 (2003).
\bibitem{schmidt2003} F.~Schmidt-Kaler, H.~Haffner, M.~Riebe, S.~Gulde,  G.P.T.~Lancaster, 
         T.~Deuschle, C.~Becher, C.F.~Roos, J.~Eschner, and R.~Blatt, 
         {\it Realization of the Cirac–Zoller controlled-NOT quantum gate},
         Nature {\bf 422}, 408 (2003).
\bibitem{roos2004} C.F.~Roos, M.~Riebe, H.~Haffner, W.~Hansel, 
         J.~Benhelm, G.P.T.~Lancaster, 
         C.~Becher, F.~Schmidt-Kaler, and R.~Blatt, 
         {\it Control and measurement of three-qubit entangled states}
         Science {\bf 304}, 1478 (2004).
\bibitem{blatt2008} R.~Blatt and D.~Wineland, 
         {\it Entangled states of trapped atomic ions},
         Nature {\bf 453}, 1008 (2008).
\bibitem{wunderlich} M.~Johanning, A.F.~Varon and C.~Wunderlich,
         {\it Quantum simulations with cold trapped ions},
          J, Phys. B: At. Mol. Opt. Phys. {\bf 42}, 154009 (2009).
\bibitem{blatt2012} R.~Blatt and C.F.~Roos,
         {\it Quantum simulations with trapped ions},
          Nature Phys. {\bf 8}, 277 (2012).
\bibitem{wineland2013} D.J.~Wineland,
          {\it Nobel Lecture: Superposition, entanglement, and raising Schrodinger's cat},
            Rev. Mod. Phys. {\bf 85}, 1103 (2013).
\bibitem{ionrevmit} C.D.~Bruzewicz, J.~Chiaverini, R.~McConnel and J.M.~Sage,
         {\it Trapped-ion quantum computing: progress and challenges},
           arXiv:1904.04178 [quant-ph] (2019).
\bibitem{monroe2018} G.~Pagano, P.W.~Hess, H.B.~Kaplan, W.L.~Tan, P.~Richerme, P.~Becker, 
         A.~Kyprianidis, J.~Zhang, E.~Birckelbaw, M.R.~Hernandez, Y.~Wu and C.~Monroe,
         {\it Cryogenic trapped-ion system for large scale quantum simulation},
         arXiv:1802.03118 [quant-ph] (2018).
\bibitem{kim10q} Y.~Nam, J.-S~Chen, N.~C.~Pisenti, K.~Wright,  C.~Delaney, D.~Maslov, 
        K.R.~Brown, S.~Allen, J.M.~Amini, J.~Apisdorf, K.M.~Beck, A.~Blinov, V.~Chaplin, 
        M.~Chmielewski, C.~Collins, S.~Debnath, A.M.~Ducore, K.M.~Hudek, M.~Keesan, S.M.~Kreikemeier, 
        J.~Mizrahi, P.~Solomon, M.~Williams, J.D.~Wong-Campos, C.~Monroe and J.~Kim,
         {\it Ground-state energy estimation of the water molecule 
         on a trapped ion quantum computer},
         arXiv:1902.10171 [quant-ph] (2019).
\bibitem{monroe2019nat} K.A.~Landsman, C.~Figgatt, T.~Schuster, N.M.~Linke, B.~Yoshida, 
          N.Y.~Yao and C.~Monroe,
         {\it Verified quantum information scrambling},
          Nature {\bf 567}, 61 (2019).
\bibitem{11q} K.~Wright, K.M.~Beck, S.~Debnath, J.M.~Amini, Y.~Nam, N.~Grzesiak, J.-S.~Chen,  N.C.~Pisenti,  
         M.~Chmielewski, C.~Collins, K.M.~Hudek, J.~Mizrahi,  J.D.~Wong-Campos,  S.~Allen, J.~Apisdorf, 
         P.~Solomon, M.~Williams, A.M.~Ducore, A.~Blinov, S.M.~Kreikemeier, V.~Chaplin, M.~Keesan, 
         C.~Monroe and J.~Kim,
         {\it Benchmarking an 11-qubit quantum computer},
          arXiv:1903.08181 [quant-ph] (2019).
\bibitem{monroefiedtheo} O.~Shehab, K.~Landsman, Y.~Nam, D.~Zhu, N.M.~Linke, M.~Keesan, 
          R.C.~Pooser and C.~Monroe,
          {\it Toward convergence of effective field theory simulations on digital quantum computers},
           arXiv:1904.04338 [quant-ph] (2019).
\bibitem{roos2019sci} T.~Brydges, A.~Elben, P.~Jurcevic, B.~Vermersch,
           C.~Maier, B.P.~Lanyon, P.~Zoller, R.~Blatt and C.F.~Roos,
           {\it Probing Renyi entanglement entropy via randomized measurements},
           Science {\bf 364}, 260 (2019).
\bibitem{aubry} S.~Aubry, 
         {\it The twist map, the extended Frenkel-Kontorova model 
          and the devil's staircase},
          Physica D {\bf 7} (1983) 240.
\bibitem{fki2007} I.~Garcia-Mata, O.V.~Zhirov, and D.L.~Shepelyansky,
         {\it Frenkel-Kontorova model with cold trapped ions},
          Eur. Phys. J. D {\bf 41}, 325 (2007).
\bibitem{haffner2011} T.~Pruttivarasin, M.~Ramm, I.~Talukdar,
           A.~Kreuter, and H.~Haffner,
           {\it Trapped ions in optical lattices for 
           probing oscillator chain models},
           New J. Phys. {\bf 13}, 075012 (2011). 
\bibitem{vuletic2015sci} A.~Bylinskii, D.~Gangloff, and V.~Vuletic,
            {\it Tuning friction atom-by-atom in an ion-crystal simulator},
             Science {\bf 348}, 1115 (2015).
\bibitem{vuletic2016natmat} A.~Bylinskii, D.~Gangloff, I.~Countis, and V.~Vuletic, 
             {\it Observation of Aubry-type transition in finite 
              atom chains via friction}, 
              Nature Mat. {\bf 11}, 717 (2016).
\bibitem{ions2017natcom} J.Kiethe, R.~Nigmatullin, D.~Kalincev, T.~Schmirander, 
             and T.E.~Mehlstaubler, 
             {\it Probing nanofriction and Aubry-type signatures 
               in a finite self-organized system},
               Nature Comm. {\bf 8} 15364 (2017).
\bibitem{diode} M.Y.~Zakharov, D.~Demidov and D.L.~Shepelyansky,
               {\it Transport properties of a Wigner crystal in one- and 
               two-dimensional asymmetric periodic potentials: Wigner crystal diode},
               Phys. Rev. B  {\bf 99}, 155416 (2019).
\bibitem{ztfki} O.V.~Zhirov, J.~Lages, and D.L.~Shepelyansky,
               {\it Thermoelectricity of cold ions in optical lattices},
                arXiv:1901.09588[cond-mat.quant-gas] (2019).
\bibitem{ztzs} O.V.~Zhirov, and D.L.~Shepelyansky,
           {\it Thermoelectricity of Wigner crystal in a periodic potential},
           Europhys. Lett. {\bf 103}, 68008 (2013).
\bibitem{chirikov} B.~V.~Chirikov, 
         {\it A universal instability 
         of many-dimensional  oscillator systems }, 
         {\em  Phys. Rep.} {\bf 52} (1979) 263.
\bibitem{lichtenberg} A.J.Lichtenberg, M.A.Lieberman, 
         {\it Regular and chaotic dynamics}, Springer, Berlin (1992).
\bibitem{meiss} J.D.~Meiss, 
          {\it Symplectic maps,
          variational principles, and transport},
          Rev. Mod. Phys. {\bf 64(3)}, 795 (1992).
\bibitem{obraun} O.M.~Braun and Yu.S.~Kivshar, 
        {\it The Frenkel-Kontorova Model: Concepts, Methods, Applications}, 
        Springer-Verlag, Berlin (2004).
\bibitem{stmapscholar} B.~Chirikov and D.~Shepelyansky, 
         {\it Chirikov standard map}, 
         Scholarpedia {\bf 3(3)}, 3550 (2008).
\bibitem{parisi} N.~Mezard, G.~Parisi and M.A.~Virasoro,
         {\it Spin glass theory and beyond},
          World Sci., Singapore (1997).
\bibitem{molmer} A.~Sorensen, and K.~Molmer,
         {\it Quantum computation with ions in thermal motion},
          Phys. Rev. Lett. {\bf 82}, 1971 (1999).
\bibitem{zoller2000nat} J.I.~Cirac and P.~Zoller,
          {\it A scalable quantum computer
            with ions in an array of microtraps},
           Nature {\bf 404}, 579 (2000).
\bibitem{aubryandre}  S.~Aubry and G.~Andre, 
              {\it Analyticity breaking and Anderson localization in incommensurate lattices},
              Ann. Israel Phys. Soc. {\bf 3}, 133 (1980).
\bibitem{inguscio} G.~Roati,  C.~D`Errico,  L.~Fallani,  M.~Fattori,  C.~Fort,
            M.~Zaccanti,  G.~Modugno,  M.~Modugno  and  M.~Inguscio, 
            {\it Anderson localization of a non-interacting Bose–Einstein condensate}, 
            Nature {\bf 453}, 895 (2008).
\bibitem{bloch} M.~Schreiber,  S.S.~Hodgman,  P.~Bordia,  H.~Luschen,
             M.H.~Fischer,  R.~Vosk,  E.~Altman,  U.~Schneider  and I.~Bloch, 
             {\it Observation of many-body localization of interacting fermions 
             in a quasirandom optical lattice},
              Science {\bf 349}, 842 (2015).
\bibitem{morogi2016pra} T.~Fogarty,  H.~Landa,  C.~Cormick and G.~Morigi,
            {\it Optomechanical many-body cooling to the ground state using frustration},
             Phys. Rev. A {\bf 94}, 023844 (2016).
\bibitem{lin2009} G.-D.~Lin, S.-L.~Zhu, R.~Islam, K.~Kim, M.-S.~Chang, S.~Korenblit, C.~Monroe and L.-M.~Duan.
             {\it Large-scale quantum computation in an anharmonic linear ion trap},
                EPL {\bf 86}, 60004  (2009).
\bibitem{blatt2d} M.~Kumph, M.~Brownnutt and R.~Blatt,
               {\it Two-dimensional arrays of radio-frequency ion traps
               with addressable interactions},
               New J. Phys. {\bf 13}, 073043 (2011).
\bibitem{schaetz2d} F.~Hakelberg, P.~Kiefer, M.~Wittemer, U.~Warring  and T.~Schaetz,
               {\it Interference in a prototype of a two-dimensional ion trap array
                quantum simulator},
                 arXiv:1812.08552 [quant-ph] (2018).
\bibitem{schaetz} Ch.~Schneider, D.~Porras, and T.~Schaetz,
         {\it Experimental quantum simulations of
          many-body physics with trapped ions},
          Rep. Prog. Phys. {\bf 75}, 024401 (2012).
\bibitem{drewsen} T.~Lauprêtre,  R.B.~Linnet,  I.D.~Leroux,  H.~Landa, A.~Dantan, and M.~Drewsen,
               {\it Controlling the potential landscape and normal modes of ion Coulomb crystals
                   by a standing-wave optical potential},
               Phys. Rev. A {\bf 99}, 031401(R) (2019).
\bibitem{home} S.~Jain, I.~Alonso, M.~Grau, and J.P.~Home,
         {\it Quantum simulation with ions in micro-fabricated Penning traps},
         arXiv:1812.06755[quant-ph] (2018).
\bibitem{anderson1958} P.W.~Anderson,
         {\it Absence of diffusion in certain random lattices},
          Phys. Rev. {\bf 109}, 1492 (1958).
\bibitem{pastur} I.M.~Lifshits, S.A.~Gredeskul and L.A.~Pastur,
         {\it Introduction to the theory of disordered systems}, 
           Wiley-Interscience, New Jersey (1988).

\end{thebibliography}
\end{document}